\documentstyle[12pt,epsf,epsfig]{article}

\title{Dynamics of Kinks in One- and Two- Dimensional Hyperbolic Models
with Quasi-Discrete Nonlinearities}
\date{\today}

\author{\\
{\bf Horacio G. Rotstein}
\thanks{E-mail: horacio@cs.brandeis.edu},
{\bf Anatol Zhabotinsky},
{\bf Irving Epstein}
\\ 
Department of Chemistry and Volen Center for Complex Systems, \\
Brandeis University, MS 015, Waltham, MA 02454-9110, USA \\
}

\def\alf{\alpha}
\def\gam{\gamma}
\def\kap{\kappa}

\def\bet{\beta}
\def\epsl{\epsilon}
\def\eps2{\epsilon^{2}}
\def\ep4s{\epsilon^{4}}

\def\ur{\rho_{1}}

\def\inveps2{\frac{1}{\epsilon^{2}}}
\def\inv3eps{\frac{1}{\epsilon^{3}}}
\def\bnu{\hat{h}}

\def\phit{\phi_{t}}
\def\phitt{\phi_{tt}}

\def\phixx{\phi_{xx}}

\def\est{s_{t}}
\def\st2{s_{t}^{2}}
\def\esx{s_{x}}
\def\sx2{s_{x}^{2}}
\def\stt{s_{tt}}
\def\sxx{s_{xx}}

\def\sxt{s_{xt}}

\def\zst2{S_{0,t}^{2}}

\def\zsx2{S_{0,x}^{2}}

\def\ost2{S_{1,t}^{2}}

\def\osx2{S_{1,x}^{2}}

\def\hrt{\rho_{t}}
\def\rht2{\rho_{t}^{2}}

\def\rhc2{\rho_{\theta}^{2}}
\def\rhtt{\rho_{tt}}

\def\zrht{\rho_{t}^{0}} 
\def\zrht2{\rho_{t}^{2}^{0}}
\def\zrhc{\rho_{\theta}^{0}}
\def\zrhc2{\rho_{\theta}^{2}^{0}}

\def\orht{\rho_{t}^{1}}
\def\orht2{\rho_{t}^{2}^{1}}
\def\orhx{\rho_{\theta}^{1}}
\def\orhx2{\rho_{\theta}^{2}^{1}}

\def\ar2{R_{0}^{2}}

\def\art2{R_{0,t}^{2}}

\def\lap-phi{\Delta \phi}
\def\lapt-phi{\Delta \phit}

\def\xicart{1 + \sx2 - \st2} 

\def\bxicart{(\xicart)^{\frac{3}{2}}}

\def\t*{t_{*}}
\def\begeq{\begin{equation}}
\def\endeq{\end{equation}}
\def\begdis{\begin{displaymath}}
\def\enddis{\end{displaymath}}
\def\bk{\bigskip}

\def\nd{\noindent}

\begin{document}

\maketitle

\begin{abstract}
We study the evolution of fronts in the Klein-Gordon equation when the 
nonlinear term is non-homogeneous. Extending previous works on homogeneous
nonlinear terms, we describe the derivation of an equation governing the front
motion, which is strongly nonlinear, and, for the two-dimensional case, 
generalizes the damped Born-Infeld equation. We study the motion of one- and 
two-dimensional fronts, finding that the dynamics is richer than in the 
homogeneous reaction term case.
\end{abstract} 

\section{Introduction}

\nd In the last few years, partial differential equations with  
discrete nonlinearities have been used to model phenomena in 
fields ranging from physics to biology, including the study of pinning of 
dislocation motions in crystals, breathers in nonlinear crystal lattices, 
Josephson junction arrays and the biophysical description of calcium release waves \cite{kn:flakla1,kn:flawil1,kn:flomaz1,kn:fuklee1,kn:leeric1,kn:cop1,kn:bishor1,kn:mitkla1,kn:mitkla2,kn:mit1,kn:klamit1,kn:ponkei1,kn:keismi1,kn:peapon1,kn:gru1,kn:gru2,kn:kee1,kn:kee2}. Recently, the discrete one-dimensional stationary version of the Klein-Gordon equation  

\begeq
	\phitt + \gam\ \phit = D\ \phixx + \alpha\  
       	\sum_{k} \delta(x-x_{k})\ [ f(\phi) + h ],
							\label{eq:intro1}
\endeq
	
\nd with \( \gam = 0 \) has been analyzed by Flach and Kladko
\cite{kn:flakla1} (see also references therein). In (\ref{eq:intro1}) 
\( \phi \) is an order parameter, the non-negative constant \( \gam \) is
the dissipation coefficient and the positive constants \( D \) and 
\( \alpha \) are the diffusion coefficient and the amplitude of the discrete 
nonlinearity, respectively. The 
function \( f \) is a bistable function (the derivative of a
double well potential having the two equal minima); i.e., a real odd function 
with positive maximum equal to \( \phi^{\ast} \), negative minimum equal to 
\( -\phi^{\ast} \)  and precisely three zeros in the closed interval 
\( [a_{-},a_{+}] \) located at \( a_{-} \), \( a_{0} \) and \( a_{+} \). 
For simplicity and without lost of generality we will consider in our analysis 
\( a_{-} = -1 \), \( a_{0} = 0 \) and \( a_{+} = 1 \). The prototype example 
is \( f(\phi) = (\phi - \phi^{3}) / 2 \). The constant \( h \), assumed to be 
small in absolute value, specifies the difference of the potential minima of 
the system; i.e., \( f(\phi) + h \) is the derivative of 
a double well potential with one local minimum and one global minimum. Note 
that (\ref{eq:intro1}) reduces to the Klein-Gordon equation when \( \sum_{k} 
\delta(x-x_{k}) \) is replaced by a constant with appropriate rescaling. In 
\cite{kn:flakla1} a first order perturbation calculation for the heteroclinic 
orbits of the corresponding stationary kink solution for (\ref{eq:intro1}) was
presented. Kink solutions, connecting the two local minima of the double well 
potential, were also obtained
for the sine-Gordon case, \( f(\phi) = - \sin(\phi) \), and the Klein-Gordon 
case, \( f(\phi) = ( \phi - \phi^{3} ) / 2 \). Both are particular cases of 
the function \( f(\phi) \) as defined above. Note that the sine-Gordon
case is equivalent to the derivative of a double well potential in a 
restricted domain of definition. 
\bk

\nd 

\nd In this manuscript, we study the dynamics of kinks for a 
quasi-discrete version of the Klein-Gordon equation 

\begeq
	\phitt + \gam \phit = D\ \lap-phi + \alpha\ \bet(x,y)\ 
	[ f(\phi) + h ],
							\label{eq:def1}
\endeq

\nd in a bounded region \( \Omega \subset R^{n} \), \( n = 1, 2 \) with smooth
boundary \( \partial \Omega \) for Neumann boundary conditions on 
\( \partial \Omega \). Equation (\ref{eq:def1}) reduces to (\ref{eq:intro1})
when \( \beta \) is one-dimensional and \( \beta(x) = 
\sum_{k} \delta(x-x_{k}) \). Altough the analysis presented below will be 
valid for a general class of positive differentiable functions \( \beta \), we 
have in mind some particular cases which approximate a distribution of 
discrete nonlinearities for large \( \eta \), a positive constant
defined below.
\bk
 
\nd Case 1) There is a sequence of points on the real line, \( x_{k} \),
\( k = 1, \ldots, N \), with \( N \) finite or infinite, where the function 
\( \beta \) reaches a maximum,
 
\begeq
	\bet(x) = \sum_{k=1}^{N}\ e^{-\eta (x-x_{k})^{2}}.
							\label{eq:def1b1}
\endeq

\nd Case 2) There is a sequence of lines in the plane, \( y_{k} \), 
\( k = 1, \ldots, N \), with \( N \) finite or infinite, where the function 
\( \beta \), independent of \( x \), reaches a maximum,

\begeq
	\bet(x,y) = \sum_{k=1}^{N}\ e^{-\eta (y-y_{k})^{2}}.
							\label{eq:def1b2}
\endeq

\nd Case 3) There is a sequence of points in the plane, \( (x_{k},y_{j}) \), 
\( k = 1, \ldots, N \), \( j = 1, \ldots, M \) with \( N \) and \( M \) finite 
or infinite, where the function \( \beta \) reaches a maximum,

\begeq
	 \bet(x,y) = \sum_{k=1}^{N} \sum_{j=1}^{M} 
	\sigma(x-x_{k},y-y_{j};\eta), 
	\ \ \ \ \ \ \ \ \ \ {\mbox where}\ \ \ \ \ \ \ \ \ \
	\sigma(x,y;\eta) = e^{-\eta(x^{2}+y^{2})}.
							\label{eq:def1c}
\endeq

\nd Case 4) There is a sequence of circles in the plane, \( \rho = \rho_{k} \),
\( k = 1, \ldots, N \), with \( N \) finite or infinite, and where \( \rho \) 
represents the radial polar coordinate, where the function 
\( \beta \) reaches a maximum,

\begeq
	\bet(\rho) = \sum_{k=1}^{N}\ 
	e^{-\eta (\rho-\rho_{k})^{2}}.
							\label{eq:def1g}
\endeq

\nd We refer to the points \( x_{k} \) and  
\( (x_{k},y_{j}) \),\( k = 1, \ldots, N \), \( j = 1, \ldots, M \) as   
quasi-discrete (QD) sites and to the stripes \( y = y_{k} \) and circles 
\( \rho = \rho_{k} \) \( k = 1, \ldots, N \) as quasi-semi-discrete (QS) sites.
We define \( d \) to be the minimum distance between two adjacent QD or QS 
sites. Note that the function \( \beta \) can be chosen to depend not only on
the spatial variable but also on \( t \). The specific form of 
\( \beta(x,y,t) \) will depend on the particular model. One might, for example,
have the product of a spatially dependent function \( \beta(x,y) \) with a 
probabilistic time-dependent function.
\bk

\nd For (\ref{eq:def1}) we define the following dimensionless variables and 
parameters

\begeq
	\hat{x} = \frac{x}{d},\ \ \ \ \ \ \ \ \ \
	\hat{y} = \frac{y}{d},\ \ \ \ \ \ \ \ \ \
	\hat{t} = \frac{\sqrt{D}\ t}{d} 
							\label{eq:def1d}
\endeq

\nd and	

\begeq
	\epsilon = \sqrt{\frac{D}{\alpha}}\ \frac{1}{d} \ \ 
	\ \ \ \ \ \ \ \ \ \
	\hat{\gamma} = \frac{\gamma\ d}{\sqrt{D}}\ \ \ \ \ \ \ \ \ \
	\hat{\eta} = \eta\ d^{2}, \ \ \ \ \ \ \ \ \ \
	\hat{h} = \frac{h}{\epsilon}.
							\label{eq:def1e}
\endeq

\nd Substituting (\ref{eq:def1d}) and (\ref{eq:def1e}) into
(\ref{eq:def1}) and dropping the \( \hat{} \) from the variables and parameters

\begeq
	\eps2\ \phitt + \eps2\ \gamma\ \phit = \eps2\ \lap-phi + \bet(x,y)\ 
	[ f(\phi) + \epsl\ h ].
							\label{eq:def1f}
\endeq
 
\nd We will consider the case \( 0 < \epsl \ll 1 \); i.e., when
diffusion is slow, \( d \) is large or \( \alpha \) is large, and 
there is a small dissipation.
\bk 

\nd The homogeneous version of (\ref{eq:def1f}), 

\begeq
	\eps2\ \phitt + \eps2\ \gam\ \phit = \eps2\ \lap-phi + f(\phi) + h,
							\label{eq:def1a}
\endeq

\nd possesses a travelling kink solution. The point on the line 
(for \( n = 1 \)) or the set of points in the plane (for \( n = 2 \)) for 
which the order parameter \( \phi \) vanishes are called the interface or 
front of the system. For (\ref{eq:def1a}) the front moves according to 
an extended version of the Born-Infeld equation \cite{kn:neu1,kn:rotnep1}

\begeq
	( 1 - \st2 )\ \sxx + 2\ \esx\ \est\ \sxt\ - ( 1 + \sx2 )\ \stt
	- \gam\ \est\ ( \xicart )  - \bnu\ \bxicart = 0, 
						\label{eq:bornexcart}
\endeq

\nd where \( \bnu \), proportional to \( h \), will be defined later. Planar 
fronts moving according to (\ref{eq:bornexcart}) with \( \gam = \bnu = 0 \) 
(no dissipation and both phases with equal potential) move with a
constant velocity equal to their initial velocity. For other values of 
\( \gam \) or \( \bnu \), fronts move with a velocity that
asymptotically approaches \( -\bnu / (\gam^{2}+\bnu^{2})^{\frac{1}{2}}
\) as long as the initial velocity  is bounded from above by \( 1 \)
in absolute value.
Linear perturbations to these planar fronts decay, either in a
monotonic or an oscillatory way, to zero as \( t \rightarrow \infty \) 
\cite{kn:rotnep1}. Circular interfaces moving 
according to (\ref{eq:bornexcart}) with \( h > 0 \) shrink to a point in 
finite time \cite{kn:neu1,kn:rotnep1}. If \( h < 0 \), then circles shrink
to points for some initial conditions and for others they grow unboundedly.
Neu \cite{kn:neu1} showed that for \( \gam = h = 0 \), closed kinks can be 
stabilized against collapse by the appearance of short wavelength, small
amplitude waves. For the more general case, two situations are possible. 
Either linear perturbations to a circle decay and curves shrink to a point in
finite time or they are still present at the shrinkage point of the
circle. Note that Equation (\ref{eq:bornexcart}) expressed in terms of
its kinematic and geometric properties reads \cite{kn:rotnep1}

\begeq
	\frac{dv}{dt} + \gam\ v\ (1 - v^{2}) - \kap\ (1 - v^{2}) +
	\bnu\ (1 - v^{2})^{\frac{3}{2}} = 0,
						\label{eq:int1geoflowmcmem}
\endeq

\nd where \( \kappa \) is the curvature of the front and \( dv/dt \) is the 
"Lagrangian" time derivative of \( v \) which is calculated along the 
trajectory fo the interfacial point moving with the normal velocity \( v \)
\cite{kn:rotnep1}.
\bk

\nd One of the goals of this papers is to determine whether the
dynamic behavior
of kinks in (\ref{eq:def1f}) differs from its homogeneous (discrete) 
nonlinearity counterpart (\ref{eq:def1a}). For the overdamped version of 
equation (\ref{eq:def1}), which is a parabolic bistable equation, it has been 
shown that there are essential differences between the homogeneous and 
non-homogeneous (discrete) cases, in that latter exhibits propagation 
failure \cite{kn:kee1,kn:kee2,kn:mit1,kn:mitkla1,kn:roteps1}.  
\bk

\nd In Section 2 we present an equation of motion for the front in equation
(\ref{eq:def1f}), and we describe briefly the method by which it was derived.
This equation generalizes equation (\ref{eq:bornexcart}) with the strong 
nonlinearity accounting for the influence of the function 
\( \beta \) on the front motion. In Section 3 we study the evolution of 
one-dimensional fronts. We show that for \( h = 0 \) the function \( \beta \)
acts as a "potential function" for the motion of the front; i.e., a front 
initially placed between two maxima of \( \beta \) asymptotically approaches 
the intervening minimum. When \( h \neq 0 \), fronts that start between two
maxima of \( \beta \) asymptotically approach an equilibrium point determined 
by \( h \) and \( \beta \), producing a kink propagation failure. In 
Section 4 we study the evolution of two-dimensional fronts with radial 
symmetry. We find that when there is no dissipation circles can shrink to a 
point in finite time, grow unboundedly or their radius can oscillate, depending
on the initial conditions. When dissipation effects are present, the 
oscillations decay spirally or not depending of the value of \( \gamma \). The
final result is the stabilization of the circular domain of one phase inside
the other phase. Our conclusions appear in Section 5.

\section{Front Dynamics: The Equation of Motion}

\nd For (\ref{eq:def1f}) and the law of motion of the interface in two 
dimensions is given by

\begdis
	( 1 - \st2 )\ \sxx + 2\ \esx\ \est\ \sxt\ - ( 1 + \sx2 )\ \stt
	- \gam\ \est\ ( \xicart )  - 
\enddis

\begeq
	- \frac{\beta_{y}(x,s) - \beta_{x}(x,s)\ \esx + \beta_{t}(x,s) \est}{2\ \beta(x,s)}\ (\xicart)
	- \bnu\ \beta^{\frac{1}{2}}(x,s)\ \bxicart = 0,
						\label{eq:bornexcartdisc}
\endeq

\nd where \( y = s(x,t) \) is the Cartesian description of the interface and
\( \bnu \) is proportional to \( h \) as will be explained later. 
Equation (\ref{eq:bornexcartdisc}) was obtained by 
carrying out a non-rigorous but self-consistent singular perturbation 
analysis for \( \epsl \ll 1 \), treating 
the interface as a moving internal layer of width \( O(\epsilon) \). We 
focused on the dynamics of the fully developed layer, and not on the process 
by which it was generated. The method that we applied is similar to that used
in \cite{kn:rotnep1} and \cite{kn:roteps1} for the study of the evolution of 
kinks in both the nonlinear wave equation (\ref{eq:def1a}) and the Allen-Cahn 
equation with quasi-discrete sources of reaction (the overdamped version of 
(\ref{eq:def1})). The basic assumptions made were:
\bk 

\nd - For small \( \epsilon \geq 0 \) and all 
\( t \in [0,T] \), the domain \( \Omega \) can be divided into two  
open regions \( \Omega_{+}(t;\epsilon) \) and \( \Omega_{-}(t,\epsilon) \) 
by a curve \( \Gamma(t;\epsilon) \), which does not intersect 
\( \partial \Omega \). This interface, defined by
\( \Gamma(t;\epsilon) := \left\{ x \in \Omega : \phi(x,t;\epsilon) = 0 
\right\} \), is assumed to be smooth, which implies that its curvature and its 
velocity are bounded independently of \( \epsilon \).
\bk

\nd - There exists a solution \( \phi(x,t;\epsl) \) of (\ref{eq:def1}), 
defined for small \( \epsl \), for all \( x \in \Omega \) and for all 
\( t \in [0,T] \)
with an internal layer. As \( \epsl \rightarrow 0 \) this solution 
is assumed to vary continuously through the interface, taking the value
\( 1 \) when \( x \in \Omega_{+}(t;\epsilon) \), \( -1 \) when 
\( x \in \Omega_{-}(t,\epsilon) \), and varying rapidly but smoothly
through the interface. 
\bk 

\nd - The curvature of of the front is small compared to its width.
\bk

\nd As a first stage in the derivation of equation (\ref{eq:bornexcartdisc}) 
we define near the interface a new variable \( z = ( y - s) / \epsl \) which 
is \( {\cal O}(1) \) as \( \epsl \rightarrow \infty \) and then express
equation (\ref{eq:def1f}) in terms of this new variable. Next we expand
\( \phi \) and \( \beta \) asymptotically in a power series in \( \epsl \) and 
substitute these expansions into the differential equation. After equating the 
coefficients of corresponding powers of \( \epsl \), we obtain two 
equations. The first can be reduced to an equation of the type
\( \Phi^{0}_{zz} + f(\Phi^{0}) = 0 \) which has to satisfy 
\( \Phi^{0}(0) = 0 \) 
and \( \Phi^{0}(\pm 1) = \pm 1 \), giving a kink solution. Here \( \Phi^{0} \) 
represents the leading order term of the order parameter \( \phi \) in terms 
of \( z \). The second problem is a linear non-homogeneous second order ODE. 
Equation (\ref{eq:bornexcartdisc}) is obtain by applying the solvability
condition (Fredholm alternative) after defining 
\( \bnu := h\ [\Phi^{0}(+\infty) - \Phi^{0}(-\infty)]\ /\
\int_{-\infty}^{\infty} (\Phi^{0}_{z})^{2} dz \).
Note that for \( f(\phi) = (\phi - \phi^{3}) / 2 \) 
(Ginsburg-Landau theory), \( \Phi^{0}(z) = \tanh \frac{z}{2} \) and
\( \bnu = 3\ h \) whereas for \( f(\phi) = \sin \phi \) 
(sine-Gordon), \( \Phi^{0}(z) = 4\ \mbox{tan}^{-1} e^{z} - \pi \)
and \( \bnu = \frac{\pi}{4}\ h \).

\section{Front Motion in 1D}

\nd For a one-dimensional system, equation (\ref{eq:bornexcartdisc}) reads

\begeq
	\stt + \gam\ \est\ ( 1 - \st2 ) + \frac{\beta'(s)}{2\ \beta(s)}
	( 1 - \st2 ) + \bnu\ \beta^{\frac{1}{2}}(s)\ ( 1 - \st2 )^{\frac{3}{2}}
 	= 0.
						\label{eq:onedim1}
\endeq

\nd We concentrate on functions \( \bet \) of the form (\ref{eq:def1b1}), 
altough the same analysis can be done for a general differentiable function.
We define \( u = s \) and \( v = \est \) obtaining

\begeq
	\left\{ \begin{array}{l}	
			u_{t} = v, 				\\
			v_{t} = - \gam\ v\ ( 1 - v^{2} ) - 
			\frac{\beta'(u)}{2\ \beta(u)} ( 1 - v^{2} ) -
			\bnu\ \beta^{\frac{1}{2}}(u)\ 
			( 1 - v^{2} )^{\frac{3}{2}}.
						\label{eq:onedim2}
		\end{array}
	\right.
\endeq 

\nd The fixed points of (\ref{eq:onedim2}) are \( (u_{0},0) \), where 
\( u_{0} \)
satisfies \( g(u) = \beta'(u) + 2\ \bnu\ \beta^{\frac{3}{2}}(u) = 0 \). The 
trace, \( \tau \), and determinant, \( \Delta \), of matrix of the linearized
 system are
\( \tau = -\gam \) and \( \Delta = [\beta''(u_{0})\ \beta(u_{0}) - 
\beta'^{2}(u_{0})]\ /\  2\ \beta^{2}(u_{0}) + \bnu\ \beta'(u_{0}) / 
[ 2\ \beta^{\frac{1}{2}}(u_{0}) ] \), respectively. If \( \bnu = 0 \) then 
the fixed points are the maxima (unstable) and minima (stable) of 
\( \beta(u) \). Thus, a front initially between two maxima of \( \beta \) will
move and asymptotically approach the intervening minimum. When there is no
dissipation, this behavior is in contrast with the homogeneous case 
(\ref{eq:bornexcart}) where, as was pointed out in the introduction, fronts
move with a constant velocity equal to their initial velocity. In the 
non-homogeneous case, we can predict the final position of the front
from the structure of \( \beta \). 
In order to understand the behavior of \( g(u) \) as 
\( \bnu \) increases above zero we consider a function \( \beta(u)  \)
with a single peak at \( 0 \); i.e., \( \bet(u) = e^{-\eta u^{2}} \). 
This function will approximate the more general (\ref{eq:def1b1}) if 
\( \eta \gg 1 \), 
so that the influence of peaks on one another is very small. In this case
\( g(u) = - 2 e^{-\eta u^{2}} [\eta\ u - \bnu\ e^{-\frac{\eta u^{2}}{2}}] \). 
For \( \bnu = 0 \), \( g(u) \) vanishes at \( u = 0 \)  and it
is positive for \( u > 0 \) and negative for \( u < 0 \). As \( \bnu \) moves 
away from zero, \( \hat{u} \), the root of \( g(u) \), will be 
given by the solution of \(\eta\ u - \bnu\ e^{-\frac{\eta u^{2}}{2}} = 0 \), 
an equation that always has a solution.
If \( \bnu > 0 \), then \( \hat{x} > 0 \), and
\( g(u) \) is positive for \( x > \hat{x} \) and negative for 
\( x < \hat{x} \). If \( \bnu < 0 \), then \( \hat{x} < 0 \). As an 
illustration, We can see the shape of \( g(u) \) as 
\( \bnu \) increases in Figure \ref{discsites-1}. In summary, as \( \bnu \) 
increases or decreases the behavior of the front is similar to the case
\( \bnu = 0 \), in contrast to the classical homogeneous case 
(\ref{eq:bornexcart}) where, as noted in Section 1, fronts
with an initial velocity whose absolute value is bounded from above by \( 1 \),
move with a velocity that asymptotically approaches 
\( -\bnu / (\gam^{2}+\bnu^{2})^{\frac{1}{2}} \).

\section{Front Motion in 2 D}

\nd The analysis of front motion in two dimensions governed by 
(\ref{eq:bornexcartdisc}) with a function \( \beta \) of type (\ref{eq:def1b2})
reduces to the analysis of one-dimensional front motion, and we shall not 
consider this case further. For radially symmetric functions, 
\( \beta = \beta(\rho) \), and radially symmetric fronts, equation 
(\ref{eq:bornexcartdisc}) for the radial coordinate \( \rho \) of the 
front reads

\begeq
	\rhtt + ( \gam\ \hrt\ + \frac{1}{\rho} )\ ( 1 - \rht2 ) +
	\frac{\beta'(\rho)}{2\ \beta(\rho)}\ (1 - \rht2) + 
	\beta^{\frac{1}{2}}(\rho)\ \bnu\ (1 - \rht2)^{\frac{3}{2}},
							\label{eq:radsim1}
\endeq
						
\nd We define \( u = \rho \) and \( v = \hrt \) obtaining

\begeq
	\left\{ \begin{array}{l}	
		u_{t} = v,						\\
		v_{t} = - [ \gam\ v + \frac{1}{u}\ +
		\frac{\beta'(u)}{2\ \beta(u)} + 
		\bnu\ \beta^{\frac{1}{2}}(u)\ (1 - v^{2})^{\frac{1}{2}} ]
		( 1 - v^{2} ).
							\label{eq:radsim2}
		\end{array}
	\right.
\endeq

\nd The lines \( v = \pm 1 \) are trajectories of (\ref{eq:radsim2}) in the 
corresponding phase plane. They define a region \( D \) with the
property that
every curve starting in this region remains inside it for all future time.
We confine our analysis to \( u > 0 \). We analyze here the case 
\( \bnu = 0 \). The fixed points of (\ref{eq:radsim2}) are \( (u_{0},0) \), 
where \( u_{0} \) are solutions (if they exist) of 
\( 2\ \beta(u) + u\ \beta'(u) = 0 \). The trace, \( \tau \), and the 
Determinant, \( \Delta \), of the matrix of the linearized system are given
by \( \tau = -\gamma \) and \( \Delta = -1 / u_{o}^{2} + 
[\beta''(u_{0})\ \beta(u_{0}) - \beta'^{2}(u_{0})]\ /\  2\ \beta^{2}(u_{0}) \)
respectively. The simplest case is \( \beta(\rho) = e^{-\eta(\rho - 
\ur)^{2}} \) for a given \( \ur > 0 \). For this case  
\( u_{0} = \left( \eta\ \ur + \sqrt{\eta^{2} \ur^{2} + 4\ \eta}
\right) / 2\ \eta \) and \( (u_{0},0) \) is a saddle point. For 
(\ref{eq:def1b2}) with \( \eta = 50 \) and \( N = 2 \), \(\rho_{1} = 0.5 \) and
\( \rho_{2} = 1.5 \), we calculated the fixed points of (\ref{eq:radsim2}) 
using the Newton-Raphson method with a tolerance of \( 0.0001 \). They are 
\( z_{1} = 0.537228 \), \( z_{2} = 0.999165 \) and \( z_{3} = 1.513217 \).
The corresponding values of \( \Delta \) are \( \Delta(z_{1}) = -53.464832 \),
\( \Delta(z_{2}) = 1196.824463 \) and \( \Delta(z_{3}) = -50.436714
\). Then \( z_{1} \) and \( z_{3} \) are saddle points, and \( z_{2} \)
is stable. Since the 
discriminant, \( \Lambda = \tau^{2} - 4\ \Delta \) of \( z_{2} \), is 
\( -4787.297852 < 0 \), \( z_{2} \) is a neutrally stable center for 
\( \gam = 0 \), a stable spiral for \( 0 < \gam \leq \gam_{0} \) and a stable 
node for \( \gam > \gam_{0} \), where \( \gam_{0} \) is the value of \( \gam \)
for which \( \Lambda = 0 \). 
\bk

\nd For the case \( \gamma = 0 \), dividing the second equation in 
(\ref{eq:radsim2}) by the first and solving one obtains

\begeq		
	c^{2}\ u^{2}\ \beta(u) + v^{2} = 1,
						\label{eq:radsim3}
\endeq

\nd where \( c^{2} = (1 - v_{i}^{2}) / ( u_{i}^{2}\ \beta(u_{i})) \) and 
\( (u_{i},v_{i}) \) are the initial conditions. In Figure \ref{level} we 
present a graph of (\ref{eq:radsim3}) for \( \beta \) given by
(\ref{eq:def1b2}) with \( N = 2 \), \( \rho_{1} = 0.5 \),
\( \rho_{2} = 1.5 \) \( \eta = 50 \) (a) and \( \eta = 10 \) (b). We observe 
that there are shrinking trajectories, oscillatory trajectories and growing 
trajectories in contrast with the homogeneous case where all trajectories 
are shrinking trajectories given by \( c^{2}\ u^{2} + v^{2} = 1 \) with 
\( c^{2} = (1 - v_{i}^{2}) / u_{i}^{2} \) \cite{kn:rotnep1}. As \( \eta \)
decreases, the oscillatory trajectories dissapear, leaving growing
trajectories, which will ultimately vanish as \( \eta \rightarrow 0 \). 
\bk

\nd In order to study more generally the behavior of the 
system away from the fixed points we can see in Figure \ref{phase-plane} the 
phase plane for \( \gamma = 0 \) (a) and \( \gamma = 1 \) (b). The dashed 
lines are the nullclines of the system and the "o" are its fixed points. The
trajectories were calculated solving (\ref{eq:radsim2}) using a Runge-Kutta
method of order four. We observe there that there are different situations
according to the initial conditions. In Figure \ref{phase-plane}-a 
(no dissipation), trajectories \( A \), \( B \) and \( C \)
correspond to circles that shrink to a point in finite time. If their initial
velocity is positive, then their radius grows initially to a
value bounded by \( z_{1} \) before shrinkage takes place. Trajectories \( D \)
and \( J \) also correspond to a circles that finally shrink to a point in 
finite time. In the case of \( D \), although the initial conditions are close 
to those of trajectory \( C \), the dynamics is very different. In
addition to shrinkage, trajectories can display unbounded growth,
represented by trajectory \( G \), and periodic behaviour, represented by 
trajectories \( E \) and \( F \). Trajectories \( H \) and \( I \) also 
correspond to circles growing unboundedly, but if the initial velocity is 
negative they shrink to a valued bounded from below by \( z_{3} \) and 
then
they start growing. In Figure \ref{phase-plane}-b (\( \gamma = 1 \))
we see that trajectories \( A \), \( B \) and \( C \) correspond to circles 
that shrink to points in finite time after growing to a radius bounded by
\( z_{1} \). Trajectory \( D \) also shrinks to a point in finite time, but
it grows initially to a radius bounded from above by \( z_{3} \) and from
below by \( z_{2} \). As we pointed out before, as a consequence of 
dissipation 
(\(\gamma \neq 0 \)) \( z_{2} \) is a stable spiral. We see that trajectory 
\( E \) spirals into \( z_{2} \), and there are no longer periodic trajectories.
For \( N > 2 \) we expect the phase plane analysis to be similar to that 
presented here. In contrast with the homogeneous nonlinear wave equation, 
where any circular front shrinks to a point in finite time, the nonhomogeneous 
version (\ref{eq:bornexcartdisc}) presents a very rich dynamics with periodic
motion and stabilization of circular domains of one phase inside the other.
\bk

\nd In the absense of dissipation there are two "forces" responsible for the 
motion of the front: the curvature of the circular front, \( 1 / \rho \), and 
the "potential function" \( \beta \). For initial conditions near
enough to the 
minimum of \( \beta \) the two "forces" balance and oscillations are possible.
When dissipation is present that "balance" is lost, and the oscillations decay.

\section{Conclusions}

\nd In this manuscript we have presented equation (\ref{eq:bornexcartdisc})
as governing the evolution of a fully developed front in a nonhomogeneous 
version of the nonlinear wave equation, (\ref{eq:def1f}), when 
\( \epsl \ll 1 \). This equation generalizes the damped version of 
the Born-Infeld equation (\ref{eq:bornexcart}) to include the effects of 
stronger nonlinearities and accounts for the influence of the 
nonhomogeneous nonlinear term on the motion of the front. The motion of 
interfaces according to (\ref{eq:bornexcartdisc}) is qualitatively different
from that of the homogeneous counterpart given by 
(\ref{eq:bornexcart}). This difference arises primarily from the fact that
the function \( \beta \) acts as a "potential function" for the motion of the
front. For the one dimensional case, an initial front initially placed 
between two maxima of \( \beta \) (which for a homogeneous nonlinear 
term will move with a velocity that  asymptotically approaches 
\( -\bnu / (\gam^{2}+\bnu^{2})^{\frac{1}{2}} \) as long as the initial 
velocity bounded from above by \( 1 \) in absolute value asymptotically 
approaches a point depending on \( \bnu \) and on the structure of \(
\beta \). For the
radially symmetric two-dimensional case, the dynamics is richer than in the 
homogeneous counterpart, where for \( \bnu = 0 \) circles 
shrink to point in finite time. In the absence of dissipation, circles can
shrink to a point in finite time, grow unboundedly or their radius  
oscillates, depending on the initial conditions. When dissipation effects are 
present, the oscillations decay, spirally or not, depending on the value of 
\( \gamma \). The final result is the stabilization of a circular domain of
one phase inside the other phase.
\bk

\nd The evolution of circular interfaces in more complicated arrangement of 
QD sites and the evolution of more complicated fronts like convex closed curves
calls for further research. We hope to address this questions in a forthcoming 
paper. 
\bk 

\nd \underline{Acknowledgement}: We thank the Chemistry Division of
the National Science Foundation for support of this work.

\bibliographystyle{unsrt}
\bibliography{tesis}

\newpage

\nd \underline{Captions}
\bk  

\begin{figure}[ph]
\begin{tabular}{llllllll}
\large \bf (a) \rm \normalsize			&
\epsfig{file=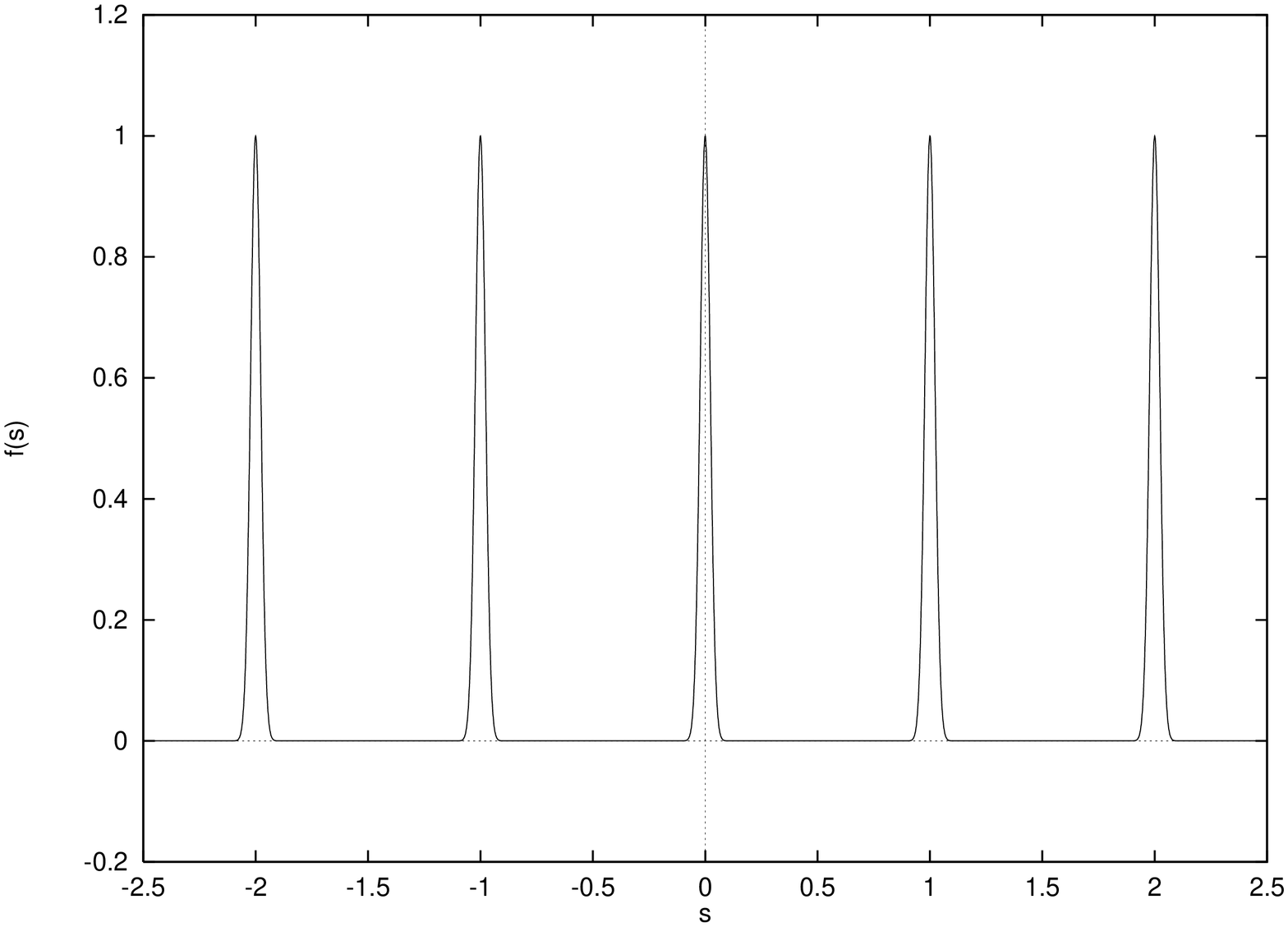,height=10cm,width=4.5cm,angle=-90}  \\
\large \bf (b) \rm \normalsize			&	
\epsfig{file=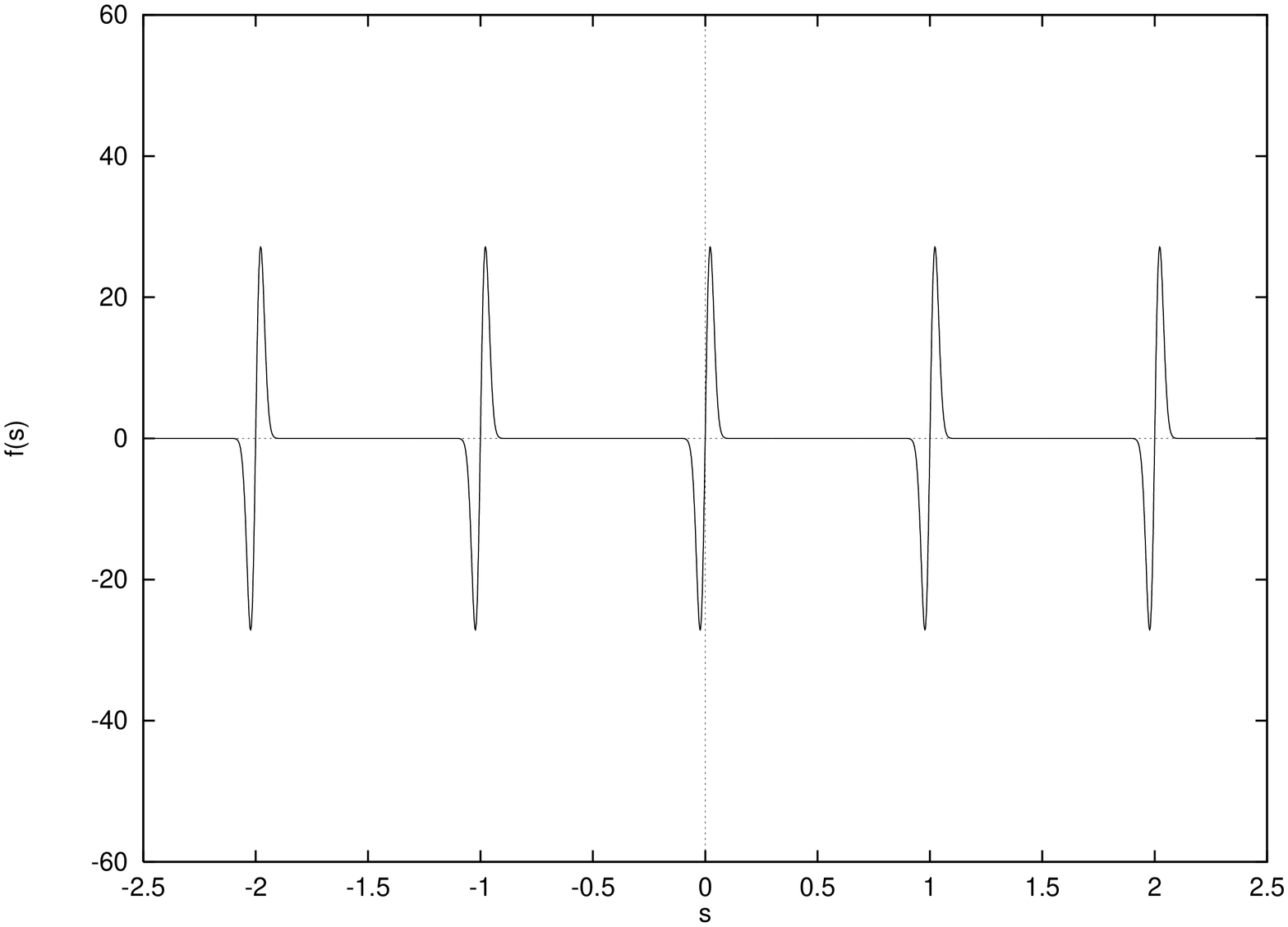,height=10cm,width=4.5cm,angle=-90}  \\
\large \bf (c) \rm \normalsize			&	
\epsfig{file=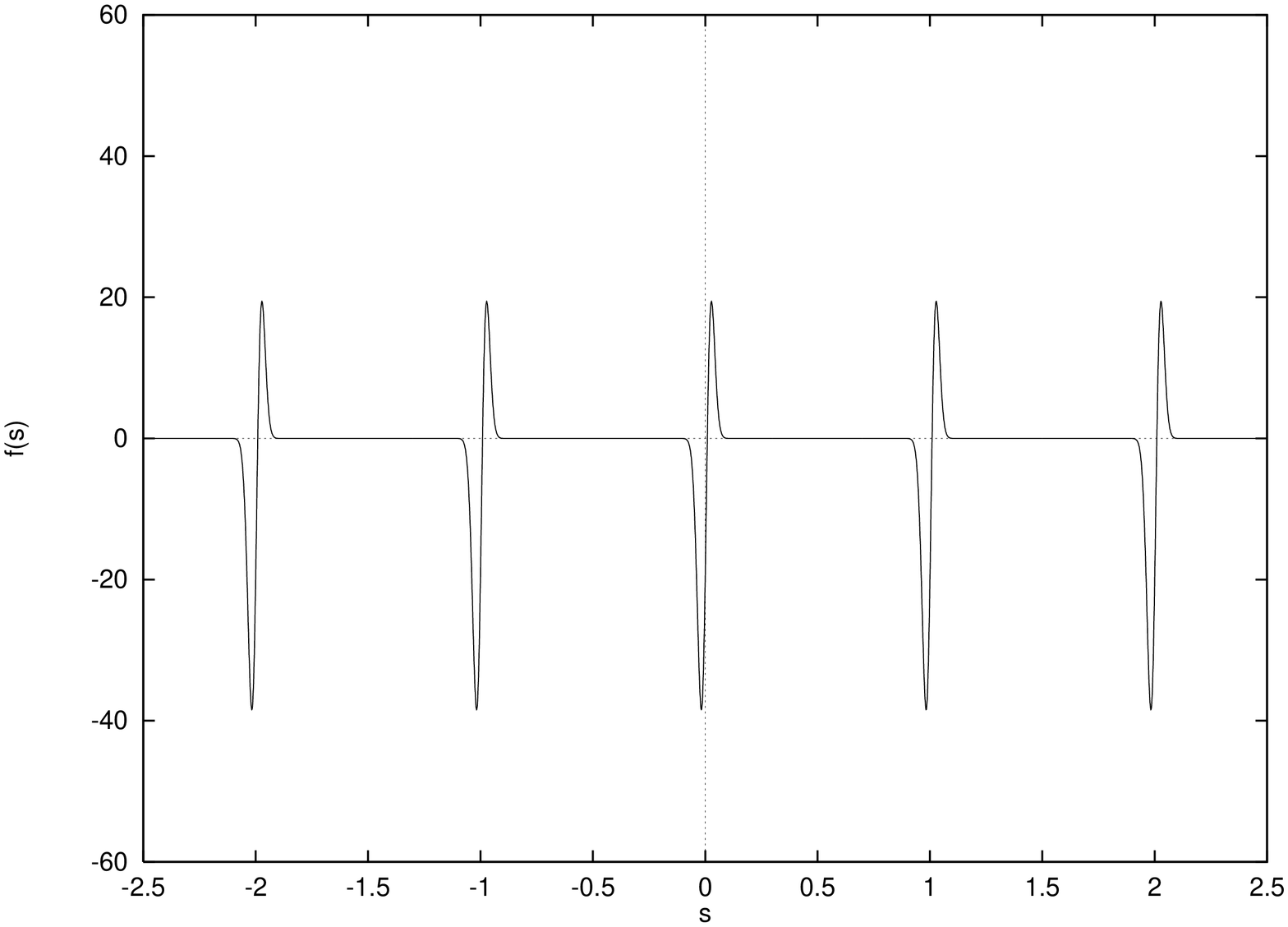,height=10cm,width=4.5cm,angle=-90}  \\
\large \bf (d) \rm \normalsize			&	
\epsfig{file=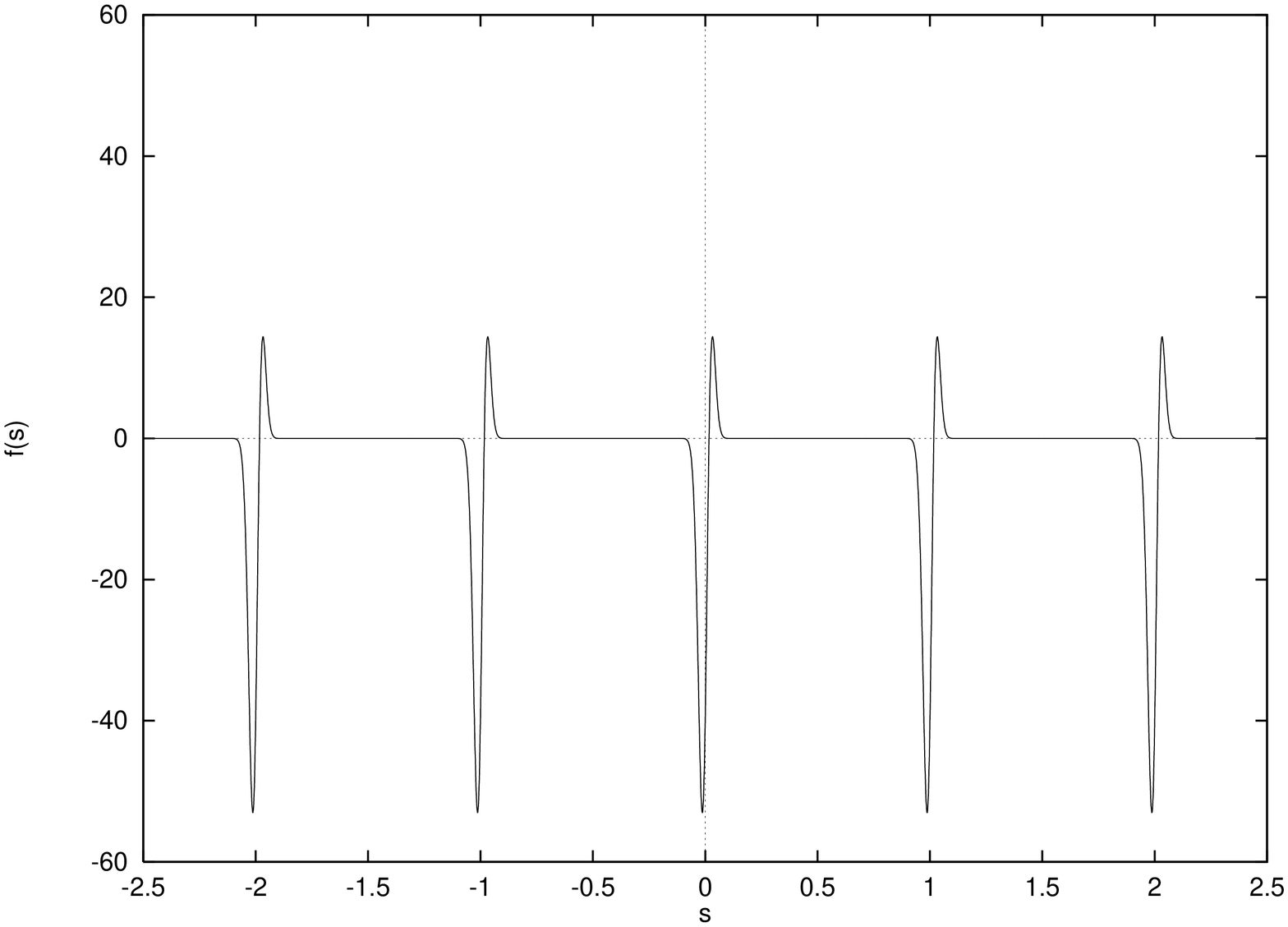,height=10cm,width=4.5cm,angle=-90}  
\end{tabular}
\caption{}
\label{discsites-1}
\end{figure}

\nd Figure \ref{discsites-1}: 

\nd a) Graph of \( \beta(u) \) for \( \eta = 1000 \), \( x_{1} = 2 \),
\( x_{2} = 1 \), \( x_{3} = 0 \) and \( x_{4} = -1 \), \( x_{5} = -2
\).

\nd b) Graph of \( g(u) \) for \( \eta = 1000 \), \( h = 0 \), \( x_{1} = 2 \),
\( x_{2} = 1 \), \( x_{3} = 0 \) and \( x_{4} = -1 \), \( x_{5} = -2
\).

\nd c) Graph of \( g(u) \) for \( \eta = 1000 \), \( h = 10 \), 
\( x_{1} = 2 \), \( x_{2} = 1 \), \( x_{3} = 0 \) and \( x_{4} = -1
\), \( x_{5} = -2 \).

\nd d) Graph of \( g(u) \) for \( \eta = 1000 \), \( h = 20 \), 
\( x_{1} = 2 \), \( x_{2} = 1 \), \( x_{3} = 0 \) and \( x_{4} = -1
\), \( x_{5} = -2 \).

\nd The points \( x_{k} \) are the maxima of \( \beta(u) \).
\bk

\begin{figure}[ph]
\begin{tabular}{llllllll}
\large \bf (a) \rm \normalsize			&
\epsfig{file=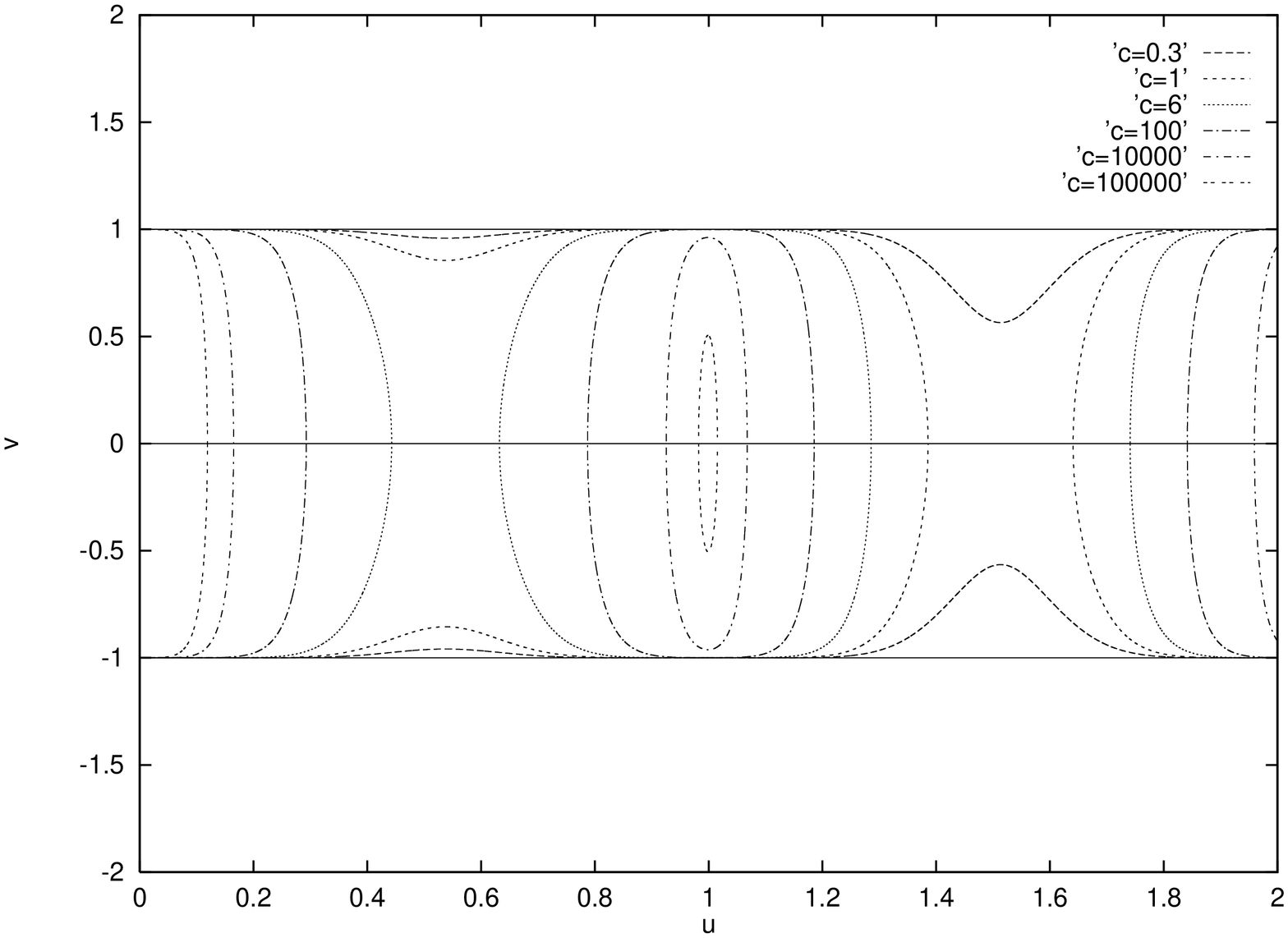,height=15cm,width=8cm,angle=-90}  \\
\large \bf (b) \rm \normalsize			&	
\epsfig{file=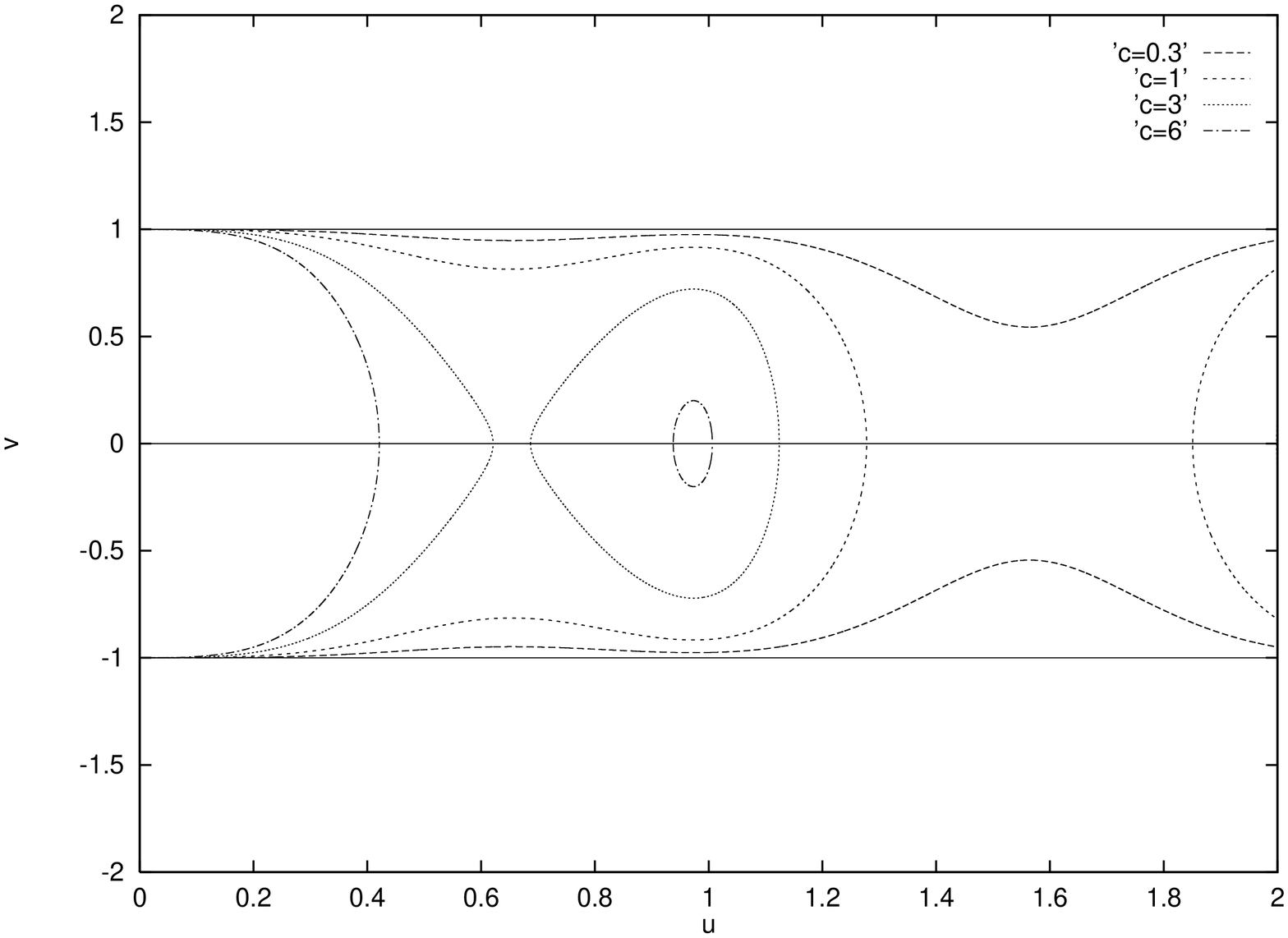,height=15cm,width=8cm,angle=-90}  
\end{tabular}
\caption{}
\label{level}
\end{figure}

\nd Figure \ref{level}:

\nd Graph of (\ref{eq:radsim3}) with \( \beta \) given by  
(\ref{eq:def1b2}) with \( N = 2 \), \( \rho_{1} = 0.5 \),
\( \rho_{2} = 1.5 \) and

\nd a) \( \eta = 50 \).

\nd b) \( \eta = 10 \).
\bk

\begin{figure}[ph]
\begin{tabular}{llllllll}
\large \bf (a) \rm \normalsize			&
\epsfig{file=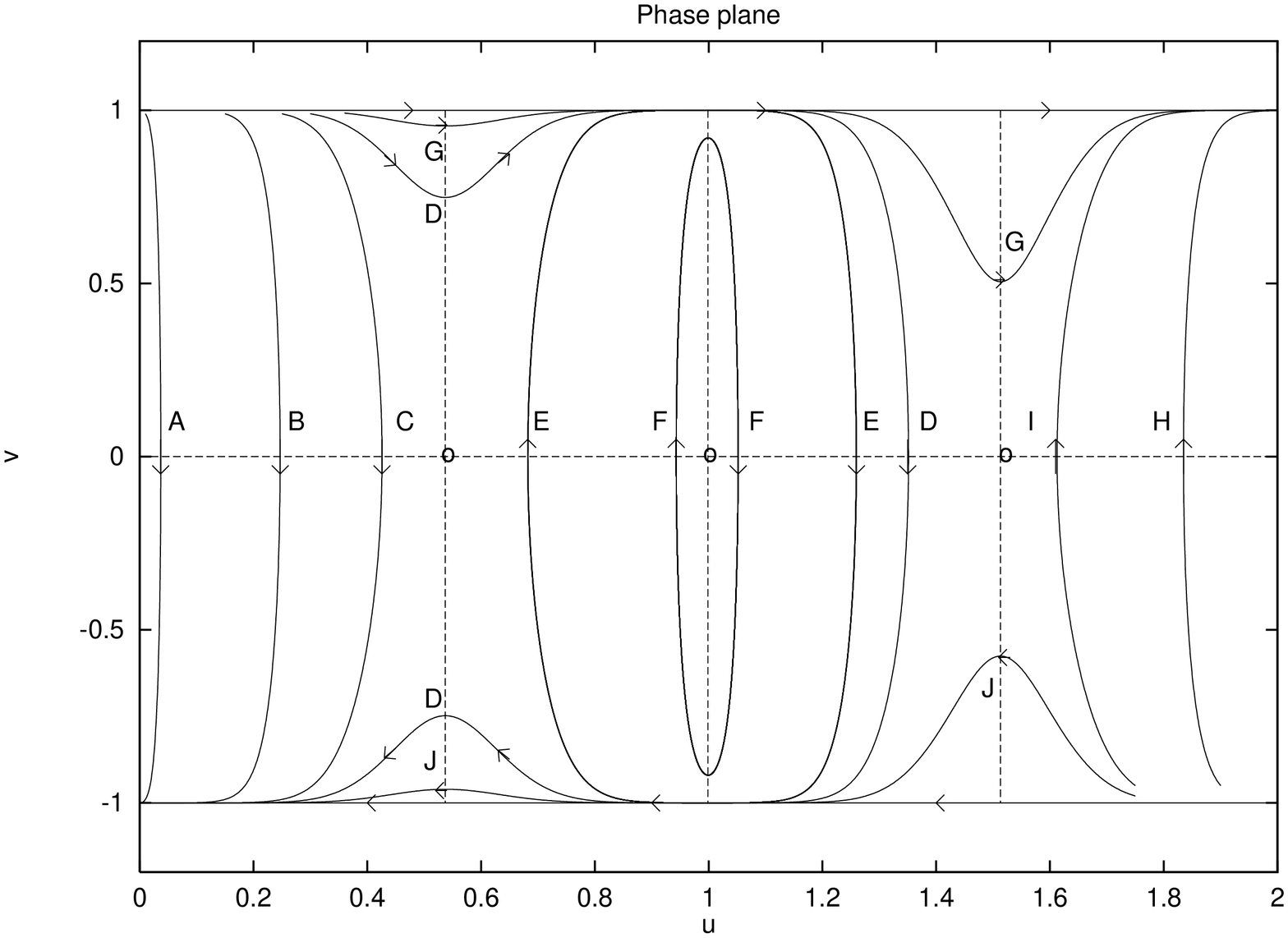,height=15cm,width=8cm,angle=-90}  \\
\large \bf (b) \rm \normalsize			&	
\epsfig{file=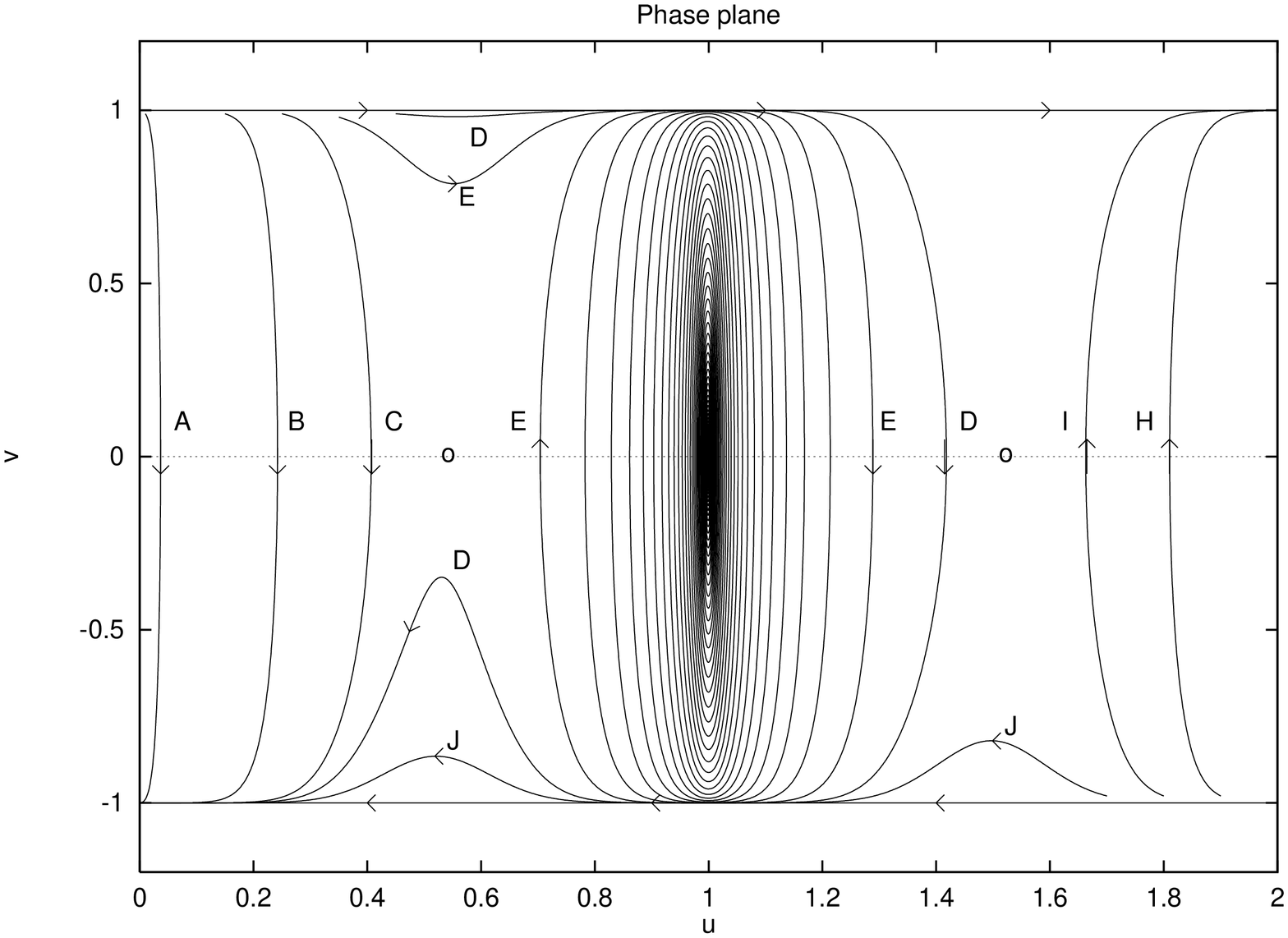,height=15cm,width=8cm,angle=-90}  
\end{tabular}
\caption{}
\label{phase-plane}
\end{figure}

\nd Figure \ref{phase-plane}: 

\nd Phase plane for (\ref{eq:radsim2}) with \( \beta \) given by 
(\ref{eq:def1b2}) with \( N = 2 \), \( \rho_{1} = 0.5 \), 
\( \rho_{2} = 1.5 \), \( \eta = 50 \) and \( \bnu = 0 \). The dashed lines
are the nullclines of the system and the "o" are its fixed points.

\nd a) \( \gamma = 0 \).

\nd b) \( \gamma = 1 \).

\newpage

\end{document}